\documentclass[aps,notitlepage]{revtex4-1}
\usepackage{amsmath,amssymb,graphicx,hyperref}
\usepackage[british]{babel}

\newcommand{\mathd}{\mathrm{d}}
\newcommand{\nobracket}{}
\newcommand{\nocomma}{}
\newcommand{\noplus}{}
\newcommand{\nosymbol}{}
\newcommand{\tmem}[1]{{\em #1\/}}
\newcommand{\tmop}[1]{\ensuremath{\operatorname{#1}}}
\newcommand{\tmstrong}[1]{\textbf{#1}}
\newcommand{\tmtextbf}[1]{{\bfseries{#1}}}

\newcommand{\keepcase}[1]{a}

\begin{document}

\title{Self-Consistent Dynamics of a Josephson Junction in the Presence of an
Arbitrary Environment\\
}

\author{Philippe Joyez}
\affiliation{Quantronics Group, Service de Physique de \\
l'Etat Condens{\'e} (CNRS URA 2464), IRAMIS,\\
CEA-Saclay, 91191 Gif-sur-Yvette, France}

\date{May 28, 2013}

\begin{abstract}
  We derive microscopically the dynamics associated with the d.c. Josephson
  effect in a superconducting tunnel junction interacting with an arbitrary
  electromagnetic environment. To do so, we extend to superconducting
  junctions the so-called $P ( E )$ theory (see e.g. Ingold and Nazarov,
  \href{http://arXiv.org/abs/cond-mat/0508728}{arXiv:cond-mat/0508728}) that
  accurately describes the interaction of a nonsuperconducting tunnel junction
  with its environment. We show the dynamics of this system is described by a
  small set of coupled correlation functions that take into account both
  Cooper pair and quasiparticle tunneling. When the phase fluctuations are
  small the problem is fully solved self-consistently, using and providing the
  exact linear admittance $Y ( \omega )$ of the interacting junction.

\end{abstract}
{\maketitle}

Fifty years ago Josephson stunned the community when he published
{\cite{josephson_possible_1962}} the equations that govern the behavior of
superconducting tunnel junctions. These Josephson relations, as they became
known, link the voltage $V$ and the superconducting phase difference $\varphi$
across the junction, and the current $I$ through it:
\begin{eqnarray}
  I=I_{0 } \sin \varphi & , & V= \frac{\hbar}{2e}  \frac{d \varphi}{dt} . 
  \label{HJ}
\end{eqnarray}
If $\varphi$ is static, $V=0$, and a nondissipative current $I$ flows through
the junction, bounded by $| I | \leqslant I_{0}$. This maximum supercurrent
$I_{0}$ (or the corresponding Josephson coupling energy $E_{J} =I_{0}   \hbar
/2e$) was originally predicted to be an intrinsic property of the tunnel
junction, depending only on its resistance in the normal state and the
superconducting gap of its electrodes {\cite{ambegaokar_tunneling_1963}}, but
not on other details such as the junction's geometry, or its fabrication
process. Along the years, Josephson junctions (JJs) have proved invaluable
electronic components forming exquisitely sensitive sensors (e.g., squid
magnetometers, quantum-limited amplifiers), metrological Volt standard
devices, or quantum bits and gates.

It is important to note that the first Josephson relation was derived assuming
that the phase $\varphi$ has negligible quantum fluctuations, and it is not
obvious why it would be generally valid beyond this situation. Because the
Josephson effect has, \ among others, metrological applications, the effect of
phase fluctuations on Josephson tunneling were thoroughly investigated in the
1980s, mostly using path integral formalism
{\cite{ambegaokar_quantum_1982,schon_quantum_1990,caldeira_quantum_1983}}. It
was concluded that in most practical experimental situations a JJ can indeed
be described using the effective Josephson Hamiltonian $H_{J} =-E_{J} \cos
\varphi$ that directly corresponds to the first Josephson relation, with,
however, small corrections due to phase fluctuations that originate in its
{\tmem{electromagnetic environment}} (i.e., the circuit connected to the
junction). This was checked for instance in the so-called Macroscopic Quantum
Tunneling experiments
{\cite{schon_quantum_1990,caldeira_quantum_1983,clarke_quantum_1988}}. More
recently, JJ-based quantum logic circuits were also shown to be accurately
described using the effective Josephson Hamiltonian
{\cite{wendin_quantum_2007}}, with their electromagnetic environment partly
responsible for their decoherence {\cite{ithier_decoherence_2005}}. Note,
however, that some environmental decoherence mechanisms in JJ qubits were
recently identified that cannot be captured within only the effective
Josephson Hamiltonian model
{\cite{martinis_energy_2009,catelani_quasiparticle_2011,ansari_effect_2012}}.

On the other hand, the environment of a JJ can have a more dramatic effect:
the phase fluctuations generated by an impedance larger than the resistance
quantum $R_{Q} =h/4e^{2} \sim 6.5  \text{k} \Omega$ are expected to suppress
the superconducting character of a JJ {\cite{schon_quantum_1990}}, and some
experiments have confirmed this prediction {\cite{corlevi_phase-charge_2006}}.
Presently several groups are actively developing nondissipative high impedance
environments using 1D arrays of JJs in the search for coherent quantum phase
slips {\cite{pop_measurement_2010,manucharyan_evidence_2012}}, or to achieve
engineering of quantum phase fluctuations
{\cite{masluk_microwave_2012,bell_quantum_2012}}. Given the goal, it is
questionable whether using the effective Josephson Hamiltonian is still fully
relevant to model these arrays. Moreover, such JJ arrays implement impedances
having several plasma mode resonances which are not readily handled by the
available theory.

In this Letter we provide a general derivation of the Josephson coupling in
the presence of phase fluctuations generated by an arbitrary electromagnetic
environment. Our derivation starts from a microscopic description of the
tunneling of individual electrons between the superconducting electrodes, and
applies the machinery of the so-called $P(E)$ theory (PoET)
{\cite{devoret_effect_1990,girvin_quantum_1990,ingold_charge_2005}}. This
theory was developed in the 1990s to explain a reduction of differential
conductance at low voltage (also called ``zero-bias anomaly'') in
nonsuperconducting sub-$\mu$m tunnel junctions, a phenomenon that is now often
referred to as dynamical Coulomb blockade. In its original form this theory
evaluates the incoherent tunneling rate of electrons properly taking into
account the probability $P ( E )$ that the environment absorbs an energy $E$
during a tunnel event. While perturbative in tunneling, this theory is
nonperturbative in the strength of the coupling to the environment and it can
deal with an arbitrary frequency-dependent linear electromagnetic environment.
Note that it also applies to incoherent Cooper pair tunneling in JJs at finite
sub-gap voltages. Its predictions were shown to be quantitative in a number of
experiments, in particular when the environment consists of resonators
{\cite{holst_effect_1994,hofheinz_bright_2011}}. Here, by generalizing PoET to
the dc Josephson effect, a coherent flow of Cooper pairs through the junction,
we obtain a unified nonperturbative treatment of arbitrary environmental
effects in both normal and superconducting tunnel junctions. In this approach
we show that one is lead naturally to introduce a self-consistent mean-field
electrodynamic response of the junction, something that, as far as we know,
has not been done explicitly previously for JJs. In this formulation the
junction is systematically and properly combined with the rest of the circuit,
resulting in an intuitive picture of the system. In the case when the phase
fluctuations are small we work out the linear response of the junction and a
simple iterative scheme to evaluate a renormalized $I_{0}$ and its admittance.
As an illustrative example, we work out the self-consistency for a JJ in an
Ohmic environment at zero temperature. In the conclusion we discuss the scope
of our results and possible extensions.

\begin{figure}[h]
  \includegraphics{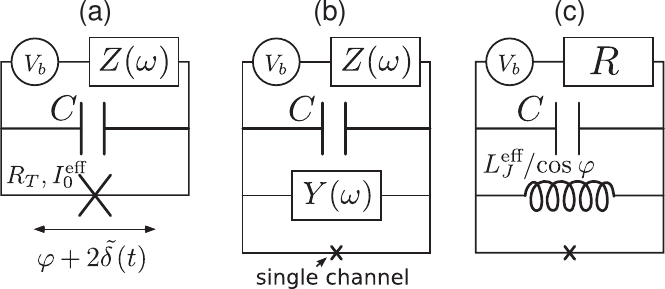}
  \caption{(a) We consider a Josephson junction characterized by its normal
  state resistance $R_{T}$ and derive its effective critical current
  $I_{0}^{\tmop{eff}}$ taking into account both a static ($\varphi$) and a
  fluctuating phase difference $\tilde{\delta} ( t )$ driven by the
  electromagnetic environment. (b) As seen from an individual tunnel channel,
  the environment consists of the impedance $Z ( \omega )$ of the connecting
  circuit, of the junction's own capacitance $C$ and of the electromagnetic
  response due to tunneling in the other channels, here described by a linear
  admittance $Y ( \omega )$, but which in the general case is a nonlinear
  element. (c) We solve the problem in the case of an Ohmic environment,
  retaining only the dominant inductive contribution in $Y.$}
\end{figure}

The circuit we consider, shown in Fig. 1a, consists of a pure tunnel element
connected in parallel with the junction's geometric capacitor and in series
with an arbitrary linear electromagnetic environment with impedance $Z (
\omega )$. The Hamiltonian of the circuit is
\[ H=H_{L} +H_{R} +H_{\text{\tmop{env}}} +H_{T} \]
where $H_{\tmop{env}}$ describes the voltage source and $Z ( \omega )$ in the
manner of Caldeira and Legget {\cite{caldeira_quantum_1983}} and $H_{L,R}$ are
the BCS Hamiltonians of the junction's electrodes. For the left electrode, for
instance, we have
\begin{eqnarray*}
  H_{L} & = & \sum_{\ell \sigma} \xi_{\ell}  c_{\ell \sigma}^{+}  c_{\ell
  \sigma}^{\nosymbol} - \Delta \sum_{\ell}  c_{\ell \uparrow}^{+} 
  c_{\bar{\ell} \downarrow}^{+} +c_{\bar{\ell} \downarrow}  c_{\ell \uparrow}
\end{eqnarray*}
where $\sigma$ is the spin index, $\ell$ is a composite channel and momentum
index for the electrons in the leads and the overbar denotes the
opposite-momentum state ($H_{R}$ has the same form, with states indexed by $r$
instead of $\ell$). Finally $H_{T} = \hat{T} + \hat{T}^{\dagger}$ is the
tunneling Hamiltonian treated as a perturbation, where the operator $\hat{T}
=e^{i  \hat{\delta}}   \sum_{\ell ,r, \sigma  } t_{\ell r}  c_{r \sigma}^{+} 
c_{\ell \sigma  }$ transfers an electron from the left to the right electrode.
We work in a gauge where the electrodes have real BCS order parameters
$\Delta$ (assumed identical in $L$ and $R$) and, consistently, the $e^{i 
\hat{\delta}}$ term here takes care of transferring the electronic charge $e$
between the electrodes
{\cite{catelani_quasiparticle_2011,ingold_charge_2005}}. We restrict to zero
dc voltage across the junction so that $\hat{\delta} ( t ) = \varphi /2+
\tilde{\delta} ( t )$ with $\varphi$ being the superconducting phase
difference across the junction and $\tilde{\delta} ( t )$ a zero-mean
fluctuating phase operator driven by $Z ( \omega )$. By introducing the
standard Bogoliubons operators
\[ \gamma_{1k} =u_{k}  c_{k \uparrow} +v_{k}  c^{+}_{\bar{k} \downarrow}   ; 
   \gamma_{0k} =-v_{k}   c_{k \uparrow} +u_{k}  c^{+}_{\bar{k} \downarrow} \]
($k= \ell ,r$) with the usual BCS coherence factors $u_{k}$, $v_{k}$ we can
diagonalize $H_{L,R}$ , whereas $H_{T}$ becomes
\begin{eqnarray*}
  H_{T} & = & \sum_{\ell ,r} t_{\ell r} [ \gamma_{0 r}^{+} \gamma_{1  \ell} (
  -e^{i  \hat{\delta}} u_{\ell} v_{r} -e^{-i  \hat{\delta}} u_{r} v_{\ell} )
  \nobracket\\
  &  & + \gamma_{1 r}^{+} \gamma_{0  \ell} ( -e^{-i  \hat{\delta}} u_{\ell}
  v_{r} -e^{i  \hat{\delta}} u_{r} v_{\ell} )\\
  &  & + \gamma_{1 r}^{+} \gamma_{1  \ell} ( e^{i  \hat{\delta}} u_{r}
  u_{\ell} -e^{-i  \hat{\delta}} v_{r} v_{\ell} )\\
  &  & \nobracket + \gamma_{0 r}^{+} \gamma_{0  \ell}   ( -e^{-i 
  \hat{\delta}} u_{r} u_{\ell  } +e^{i  \hat{\delta}} v_{r} v_{\ell} ) ] + (
  \ell \leftrightharpoons r )^{\dag} .
\end{eqnarray*}
In thermal equilibrium situations the supercurrent through the junction is
given by the thermodynamic relation
\begin{equation}
  I= \frac{2e}{\hbar} \frac{dF}{d \varphi} \label{IS}
\end{equation}
where $F$ is the free energy. To lowest order in perturbation theory the
change of $F$ due to $H_{T}$ can be cast as
\begin{equation}
  \Delta F= \frac{1}{\hbar} \int_{0}^{+ \infty}  d t    \tmop{Im}  S_{H_{T}} (
  t ) \label{deltaF}
\end{equation}
with $S_{H_{T}} ( t ) = \langle H_{T} ( t ) H_{T} ( 0 ) \rangle$ where the
angular brackets denote averaging over the unperturbed quasiparticle and
environment states that act as bath degrees of freedom whose time evolution is
the unperturbed one. A straightforward algebraic calculation gives
\begin{eqnarray}
  S_{H_{T}} ( t ) & = & \sum_{\ell ,r, \eta = \pm} | t_{\ell  r} |^{2} [ 
  u_{r} v_{r} u_{\ell} v_{\ell} (A_{\eta} ( t ) -B_{\eta} ( t ) )C_{\eta \eta}
  ( t )  e^{i \eta \varphi}    \nobracket \nonumber\\
  &  &                                                                
  \nobracket  + ( ( u_{\ell}^{2} v_{r}^{2} +u_{r}^{2} v_{\ell  }^{2} )
  A_{\eta}   ( t ) +  ( u_{\ell  }^{2} u_{r}^{2} +v_{\ell  }^{2} v_{r}^{2} )
  B_{\eta} ( t ) ) C_{\eta - \eta} ( t ) ]  \label{HTcorr}
\end{eqnarray}
with
\begin{eqnarray*}
  A_{\eta = \pm} ( t ) & = & \langle \gamma_{0  \ell  }^{\eta} (t) \gamma_{0 
  \ell  }^{- \eta} \gamma_{1 r}^{- \eta} (t) \gamma_{1 r}^{\eta} \noplus +
  \gamma_{1  \ell}^{- \eta} (t) \gamma^{\eta}_{1 \ell} \gamma_{0 r }^{\eta}
  (t) \gamma_{0 r }^{- \eta} \rangle\\
  B_{\eta} ( t ) & = & \langle \gamma_{0  \ell  }^{\eta} (t)
  \gamma_{0  \ell  }^{- \eta} \gamma^{- \eta}_{0 r} (t) \gamma^{\eta}_{0 r} +
  \gamma^{- \eta}_{1  \ell} (t) \gamma_{1  \ell}^{\eta}   \gamma_{1 r }^{\eta}
  (t) \gamma^{- \eta}_{1 r } \rangle\\
  C_{\eta \eta'} ( t ) & = & \langle \text{$e^{i \eta   \tilde{\delta} ( t )}
  e^{i \eta'   \tilde{\delta} ( 0 )}$} \rangle
\end{eqnarray*}
where a fermion operator with a minus exponent means an annihilation operator.
The $e^{\pm i \varphi}$ terms in Eq. (\ref{HTcorr}) are each related to the
transfer of two spin-conjugate electrons in a given direction, i.e., a whole
Cooper pair with charge $2e$, they thus correspond to the Josephson effect.
Note also that they come with the $u_{r} v_{r}$ and $u_{\ell} v_{\ell}$
factors that correspond to the anomalous Green's function of the electrodes,
carrying the essence of superconductivity. The $\varphi$-independent terms, on
the contrary, are related to a back-and-forth transfer of an electron and
correspond to ordinary quasiparticle tunneling, the only processes remaining
in the normal state. These processes do not transfer a net charge through the
junction but they still couple to the phase fluctuations and contribute to the
dynamics of the JJ. While these processes are obviously disregarded when JJs
are modeled using only the effective Josephson Hamiltonian (e.g. most JJ-based
qubit literature), the full Ambegaokar-Eckern-Sch{\"o}n effective action for
the JJ {\cite{ambegaokar_quantum_1982}} [whose form is closely related to Eq.
(\ref{HTcorr})] allows accounting for them in path integral formalism. In the
present approach we handle these terms using only two-point real-time
correlators and sparing the use of path integrals. The correlators $C_{+ -} (
t ) , C_{-+} ( t )$ that accompany quasiparticle tunneling are those
encountered in the standard PoET [specifically, $\langle \text{$e ^{i 
\tilde{\delta} ( t )} e ^{-i  \tilde{\delta} ( 0 )}$} \rangle = \int \mathd  t
e^{-i Et/ \hbar} P ( E )$ is the inverse Fourier transform of $P ( E )$],
while the Cooper pair tunneling comes with distinct correlators $C_{+ +} ( t )
,C_{- -} ( t )$. For simplicity we here assume phase fluctuations are
symmetric, {\tmem{i.e.}} $C_{+ +} =C_{- -}$ and $C_{+ -} = C_{-+}$ (we discuss
the limit of validity of this assumption in the Supplemental Material
{\cite{supplementary_material}}). Going to a continuum of states in the
electrodes, from Eq. (\ref{HTcorr}) we obtain the exact result at lowest order
in tunneling
\begin{equation}
  S_{H_{T}} ( t ) = \frac{2R_{Q}}{\pi^{2} R_{T}} [ ( p ( t )^{2} -q ( t )^{2}
  ) C_{+-} ( t ) +m ( t )^{2} C_{++} ( t )   \cos   \varphi ] \label{HTHT}
\end{equation}
where $R_{T}$ is the normal state tunnel resistance of the junction and $m ( t
) ,p ( t ) \nocomma \nocomma , q ( t )$ are, respectively, the inverse Fourier
transforms of $\mathcal{M} ( \varepsilon )  = - \Delta f ( - \varepsilon )
\rho ( \varepsilon ) / \varepsilon$, $ \mathcal{P} ( \varepsilon )  = f ( -
\varepsilon ) \rho ( \varepsilon )$, $\mathcal{Q} ( \varepsilon )  =-f ( -
\varepsilon ) \theta \left( \varepsilon^{2} - \Delta  ^{2} \right) \tmop{sgn}
( \varepsilon )$ with $\rho ( \varepsilon ) = | \varepsilon |   \tmop{Re}
\left( \varepsilon^{2} - \Delta  ^{2} \right)^{-1/2}$ the BCS density of
states, $\theta$ the Heavyside step and $f ( \varepsilon )$ the occupation
probability of the Bogoliubov quasiparticles, which need not be thermal. Here
both electrodes are assumed identical but the general case could also be
handled. Note that in principle the gap $\Delta$ of the electrodes should be
self-consistently evaluated from $f ( \varepsilon )$, an effect which becomes
important at temperatures comparable to the critical temperature or in strong
nonequilibrium. If we first ignore a possible $\varphi$ dependence of
$C_{\noplus \noplus + \pm}$, then, by combining Eqs. (\ref{IS}),
(\ref{deltaF}), and (\ref{HTHT}) one obtains a generalization of the first
Josephson relation with an effective critical current
\begin{equation}
  I_{0}^{\tmop{eff}} = \frac{2}{\pi e \hbar R_{T}} \left| \int_{0}^{+ \infty} 
  d t   \tmop{Im}   [ m ( t )^{2} C_{++} ( t ) ] \right| \label{I0eff1}
\end{equation}
which remains valid beyond thermal equilibrium. This expression generalizes
PoET in real-time formulation
{\cite{joyez_single-electron_1997,odintsov_effect_1988}}. In the case where
phase fluctuations are negligible $C_{++} ( t ) \equiv 1$, and one recovers
all standard results on JJ, such as, {\tmem{e.g.}}, the temperature dependence
of the critical current {\cite{ambegaokar_tunneling_1963}}. Hence $C_{++}$ is
a kernel giving a renormalization of the critical current with respect to the
standard Ambegaokar-Baratoff value {\cite{ambegaokar_tunneling_1963}}. We will
see below that $C_{\noplus \noplus + \pm}$ should in principle depend on
$\varphi$ (albeit weakly in usual cases), thus yielding additional terms that
cause a departure from the purely sinusoidal current-phase relation predicted
by Josephson.

We now consider finite phase fluctuations and first assume that the degrees of
freedom generating these fluctuations can be regarded as a linear impedance
$Z_{\tmop{eff}}$ as in the usual PoET {\cite{ingold_charge_2005}}. Such
fluctuations are then Gaussian and consequently $C_{\noplus \noplus + \pm}$
can be expressed in terms of only the two-point correlator $S_{\delta} ( t ) =
\langle \tilde{\delta} ( t ) \tilde{\delta} ( 0 ) \rangle$. As a consequence
of the fluctuation-dissipation theorem $S_{\delta} ( t )$ can in turn be
evaluated from the spectral density of the environment. Namely
\begin{eqnarray}
  C_{+-} & = & e^{S_{\delta} ( t ) -S_{\delta} ( 0 )} =e^{J ( t )} \nonumber\\
  C_{+ +} & = & e^{-S_{\delta} ( t ) -S_{\delta} ( 0 )} =e^{-J ( t )
  -2S_{\delta} ( 0 )}  \label{C++}\\
  S_{\delta} ( t ) & = &  \int_{- \infty}^{+ \infty} \frac{d \omega}{\omega} 
  \frac{\text{\tmop{Re}}  Z_{\tmop{eff}} ( \omega )}{2R_{Q}}  \frac{e^{-i
  \omega t}}{1-e^{- \beta \hbar \omega}} .  \label{Sdelta}
\end{eqnarray}
Here we have also introduced the usual PoET notation $J ( t ) =S_{\delta} ( t
) -S_{\delta} ( 0 )$ {\cite{ingold_charge_2005}}. Replacing $C_{+ +}$ in Eq.
(\ref{I0eff1}) we can pull out of the integral the renormalization factor
$\lambda =e^{-2S_{\delta} ( 0 )}$, \ which plays a major role in the
following.

Note that, unless $\text{\tmop{Re}} Z_{\tmop{eff}} ( \omega \sim 0)
=\mathcal{O} ( \omega^{2} )$ or smaller, $S_{\delta} ( 0 ) = \infty$
(signaling thermal or quantum phase diffusion), yielding $\lambda =0$ and thus
$I_{0}^{\tmop{eff}} =0.$ This might seem surprising since in most cases when
one measures a JJ, it is connected to a circuit that contains normal metal at
room temperature (with finite dc resistance), but its critical current is
nevertheless measured finite. The apparent paradox is resolved when one
considers the JJ as being part of its own electromagnetic environment [see
Fig. 1(b)] : a superconducting JJ perfectly shunts the rest of the circuit at
zero frequency, preventing phase diffusion and the divergence of $S_{\delta} (
0 )$. More importantly, doing so is actually the only way to enforce an
amplitude and a dynamics of the phase fluctuations in the system that are
actually consistent with the presence of the junction, unlike in standard PoET
{\cite{note1}}. This inclusion of the junction in its own environment can also
be justified microscopically: a typical metallic tunnel junction contains a
very large number $N$ of independent Landauer channels that only interact
through their common phase. Thus, as seen from each individual channel, the
other channels form a ({\tmem{a priori}} nonlinear) bath whose response is
that of the full junction (up to corrections of order $1/N$) and which are
treated like the rest of the environment. Let us stress also that in typical
tunnel junctions even if the junction's conductance is large, its individual
channels remain very weakly transmissive. Hence, lowest order perturbation in
tunneling is sufficient and all the complications in the behavior of the JJ
arise solely from the electromagnetic interaction among the channels and with
the environment, which treat here in a self-consistent mean-field manner. Such
a self-consistent mean-field approach of PoET has been successfully checked
experimentally in low-resistance normal-state junctions
{\cite{joyez_how_1998}}, and, in that case, when the junction is described as
a linear element (see below), this was shown to correspond to a
self-consistent harmonic approximation that minimizes the free energy in the
path integral description of the system
{\cite{goppert_frequency-dependent_1999}}. Let us finally remark that in this
mean-field approach the superconducting character of the JJ gives rise to a
chicken-and-egg situation that requires a self-consistent solution, much like
for the value of $\Delta$ in BCS theory itself.

We now close the loop by working out the self-consistency in the linear
regime assumed in this part. Within this hypothesis, the response of the
junction can be obtained from a generalized fluctuation-dissipation relation
{\cite{safi_time-dependent_2011}} and is expressed as an admittance
\begin{equation}
  Y ( \omega ) = \frac{\cos   \varphi \noplus}{i L_{J }^{\tmop{eff}} \omega}
  +2 \int_{0}^{\infty} dt   i  \tmop{Im}  S_{I } ( t ) \frac{e^{i \omega t}
  -1}{\hbar \omega} \label{Y}
\end{equation}
that is exact at lowest order in perturbation {\cite{supplementary_material}}.
In this expression $L_{J}^{\tmop{eff}} = \left( \frac{2e}{\hbar}
I_{0}^{\tmop{eff}} \right)^{-1}$ is the effective Josephson inductance and
$S_{I } ( t ) = \langle \hat{I} ( t ) \hat{I} ( 0 ) \rangle$ is the correlator
of the current operator $\hat{I}$=$\frac{2e}{\hbar}  \frac{\partial 
H_{T}}{\partial   \varphi}$ through the junction. This latter definition
implies that $S_{I } ( t, \varphi ) = \left( \frac{e}{\hbar} \right)^{2}
S_{H_{T}} ( t, \varphi + \pi )$, readily obtained from Eq. (\ref{HTHT}). In
the self-consistent approach we discuss here we shall then replace
\begin{equation}
  Z_{\tmop{eff}} ( \omega ) = [ Y ( \omega ) +iC \omega +Z^{-1} ( \omega )
  ]^{-1} \label{Zeff}
\end{equation}
in Eq. (\ref{Sdelta}), where $C$ is the junction capacitance and $Z$ the
impedance of the external circuit as seen from the junction [see Fig.1(b)].
Thus we are able to obtain the full dynamics of the system (and
$I_{0}^{\tmop{eff}}$ as a by-product) by solving the self-consistency defined
by Eqs. (\ref{HTHT}), (\ref{Y}), (\ref{Zeff}), (\ref{Sdelta}), (\ref{C++}).
This can, for instance, be done by iterating from an initial guess such as $Y
( \omega ) = \cos   \varphi /i L_{J }^{0} \omega$, $L_{J}^{0}$ being the
Josephson inductance in the absence of environment. In order to be valid the
iterated solution must be consistent with the assumption of linear behavior of
the effective environment, i.e.,
\begin{equation}
  \sqrt{ S_{\delta} ( 0 )} \ll 2 \pi , \label{linearvalid}
\end{equation}
so that phase fluctuations do not feel the nonlinearity of the JJ. In practice
this means $I_{0}^{\tmop{eff}}$ should not be reduced more than a few percent
with respect to $I_{0}$ for this linear approach to be valid. If this later
criterion if fulfilled, then the solution obtained is essentially the exact
dynamics of the junction at lowest order in tunneling.

Simplifying approximations can be made or not depending on the value of the
``plasma frequency'' $\omega_{p} = ( \cos   \varphi /L_{J}^{\tmop{eff}} C
)^{1/2}$ defined as the resonance frequency of the purely inductive first term
of Eq. (\ref{Y}) with the junction's capacitance $C$. If $\omega_{p}$ is
significantly smaller than $\omega_{\tmop{Gap}} =2 \Delta / \hbar$, then at
low temperature it is a good approximation to keep in $Y ( \omega )$ only the
inductive term, that precisely suppresses the divergence of $S_{\delta} ( 0
)$. This is justified because the integral in Eq. (\ref{Y}) has only a slight
capacitive contribution at frequencies $\omega \lesssim \omega_{\tmop{Gap}} =2
\Delta / \hbar$ with dissipation setting in only at frequencies close to or
above $\omega_{\tmop{Gap}}$. With this simplification $Z_{\tmop{eff}}$ reduces
to the impedance of an LC oscillator resonating at $\omega_{p}$ damped by the
external impedance $Z ( \omega )$. Furthermore, still in the case when
$\omega_{p} < \omega_{\tmop{Gap}}$, the characteristic time scale of phase
fluctuations ($\omega_{p}^{-1}$) is significantly longer than that of $m^{2} (
t )$ which is $\omega_{\tmop{Gap}}^{-1}$. Then, in Eqs. (\ref{I0eff1}),
(\ref{C++}) we can take the short-time limit $J ( t \rightarrow 0 ) =0$,
yielding the simple renormalization $I_{0}^{\tmop{eff}} = \lambda I_{0}$. A
similar renormalization of the Josephson coupling was obtained at $\varphi =0$
in Refs. {\cite{schon_quantum_1990,grabert_phase_1998}}. We see here that this
is valid only when $\omega_{\tmop{Gap}}$ is the fastest dynamics in the
problem and that the opposite situation cannot be treated correctly in
approaches starting from the effective Josephson Hamiltonian.

Let us now fully work out an example in the above simplifying assumption
$\omega_{p} < \omega_{\tmop{Gap}}$, $I_{0}^{\tmop{eff}} = \lambda I_{0}$, and
further restricting to the ``Ohmic'' case where $Z ( \omega ) =R$ [Fig.1(c)]
and zero temperature. Then the effective environment reduces to an RLC circuit
with impedance $Z_{\tmop{eff}} ( \omega ) = ( \lambda   \cos   \varphi /i
\omega L_{J}^{0} +iC \omega +R )^{-1}$ for which $S_{\delta} ( 0 )$ can be
calculated analytically and from which we derive the self-consistency equation
\begin{equation}
  \lambda = \exp - \frac{R}{2R_{Q}}   \frac{  \tanh  ^{-1} \left( \frac{1-2 
  \lambda  q^{2} \cos   \varphi}{\sqrt{1-4  \lambda  q^{2} \cos   \varphi}}
  \right) +i  \frac{\pi}{2}  }{  \sqrt{1-4  \lambda  q^{2} \cos   \varphi}}  
  \label{selfcons}
\end{equation}
where $q=R \sqrt{C/L_{J}^{0}}$ would be the quality factor of the plasma
oscillation at $\varphi =0$, in absence of renormalization. Again, valid
solutions must satisfy Eq. (\ref{linearvalid}), that is, $1- \lambda \ll 1$.
However this always fails at $\varphi = \frac{\pi}{2}   \tmop{mod}   \pi$
where $\omega_{p}$ vanishes and where a treatment beyond linear response is
needed. When the approximation is valid (away from the pathological points) we
predict that the renormalization of $I_{0}$ is different at $\varphi =0$ and
$\varphi = \pi$, leading to a slightly anharmonic current-phase relation. This
anharmonicity is a generic feature in the self-consistent approach because it
causes $C_{+  \pm}$(t) in Eq. (\ref{HTHT}) to have a $\varphi$ dependence
through the dynamical response of the JJ.

In conclusion we have extended the framework of the PoET to address the effect
of an arbitrary electromagnetic environment on the Josephson effect in
metallic tunnel junctions. Doing so we reached a self-consistent description
of the Josephon effect, sheding new light on the interaction of a JJ with its
environment, including its dynamics. This notably predicts that the celebrated
first Josephson relation generically departs from a sinusoid when the
impedance of its environment is increased, a fact that should be verifiable
experimentally. For strictly dc Josephson effect and small phase fluctuations,
the self-consistency is fully worked out using the exact linear admittance of
the interacting JJ, a quantity that is accessible to measurements and that
should be useful for quantum circuit engineering. We think more work in this
direction could extend this approach to non-dc situations and non-Gaussian
phase fluctuations {\cite{supplementary_material}}. This would provide the
general ``circuit laws'' for Josephson junctions, a quantum nonlinear
generalization of the classical ``impedance combination laws.''

The author is grateful to all members of the Quantronics group for their
constant interest and support and thankfully acknowledges helpful discussions
and input from C. Altimiras, H. Grabert, F. Hekking, M. Hofheinz, H. le Sueur,
F. Portier, P. Roche and I. Safi.

\label{Supplemantary}\section*{Supplemental material}

\subsection*{Derivation of the JJ admittance}

Here we evaluate the linear response of the junction to a vanishingly small ac
excitation $\delta  V( \omega  )=i \omega   \frac{\hbar  }{2e} \delta  
\varphi  ( \omega  )$ added to the static phase difference $\varphi$ of the
junction. This can be done exactly, even in presence of the environment
{\cite{safi_time-dependent_2011}}. At the lowest order in the tunneling
Hamiltonian and in the excitation, the time evolution of the current flowing
through the junction under this perturbation is given by
\begin{eqnarray*}
  I(t ) & = & \frac{i}{\hbar} \int_{- \infty}^{t} \mathd  s  \langle [ \hat{I}
  ( t ) ,H_{T} ( s ) ] \rangle\\
  & = & \frac{i}{\hbar} \int_{- \infty}^{t} \mathd  s  \left\langle \left[
  \hat{I} ( t ) + \delta \varphi ( t ) \frac{\partial \hat{I}}{\partial
  \varphi} ( t ) ,H_{T} ( s ) + \delta \varphi ( s ) \frac{\partial
  H_{T}}{\partial \varphi} ( s ) \right] \right\rangle\\
  & = & \langle \hat{I} \rangle + \frac{i}{\hbar} \delta \varphi ( t )
  \int_{- \infty}^{t} \mathd  s  \left\langle \left[ \frac{\partial
  \hat{I}}{\partial \varphi} ( t-s ) ,H_{T} \right] \right\rangle +
  \frac{i}{\hbar} \int_{- \infty}^{t} \mathd  s  \left\langle \left[ \hat{I} (
  t-s ) , \frac{\partial H_{T}}{\partial \varphi} \right] \right\rangle \delta
  \varphi ( s )
\end{eqnarray*}
where, as in the body of the article, the angular brackets denote averaging
over unperturbed states of the electrode and the environment and the time
evolution of operators is the unperturbed one. $\langle \hat{I} \rangle$ is
the dc supercurrent in absence of the ac excitation. Using the identities :
\begin{eqnarray*}
  \frac{\partial  H_{T}}{\partial \varphi} = \frac{\hbar}{2e} \hat{I} &       
  ;             & \frac{\partial \widehat{ I}}{\partial \varphi} =-
  \frac{e}{2 \hbar} H_{T}  
\end{eqnarray*}
\[ \begin{array}{lll}
     S_{I } ( t, \varphi ) = \left( \frac{e}{\hbar} \right)^{2} S_{H_{T}} ( t,
     \varphi + \pi ) &  & 
   \end{array} \]
and denoting $\delta I ( t ) =I(t )- \langle \hat{I} \rangle$, and $S_{I}^{-}
( t )$ the odd part of $S_{I} ( t )$ we get
\begin{eqnarray*}
  \delta I ( t ) & = & - \frac{e}{2 \hbar} \frac{i}{\hbar}
  \frac{\hbar^{2}}{e^{2}} \delta \varphi ( t ) \int_{- \infty}^{t} \mathd s 
  2S_{I}^{-} ( t-s, \varphi + \pi ) + \frac{i}{2e} \int_{- \infty}^{t} \mathd 
  s  2S_{I}^{-} ( t-s, \varphi ) \delta \varphi ( s )\\
  & = & \frac{i}{2e} \left( - \delta \varphi ( t ) \int_{- \infty}^{\infty}
  \mathd s  2S_{I}^{-} ( s, \varphi + \pi ) \theta ( s ) + \int_{-
  \infty}^{\infty} \mathd s  2S_{I}^{-} ( t-s, \varphi ) \theta ( t-s ) \delta
  \varphi ( s ) \right)
\end{eqnarray*}
Going to the frequency domain
\begin{eqnarray*}
  \delta I ( \omega ) & = & \frac{i}{2e} \delta \varphi ( \omega ) \int \mathd
  t    \theta ( t ) ( 2S_{I}^{-} ( t, \varphi ) e^{i \omega t} -2S_{I}^{-} (
  t, \varphi + \pi ) )\\
  & = & \frac{i}{2e} \delta \varphi ( \omega ) \left(  -i2e
  I_{0}^{\tmop{eff}} \cos   \varphi + \int \mathd  t    \theta ( t )
  2S_{I}^{-} ( t, \varphi ) ( e^{i \omega t} -1 ) \right)
\end{eqnarray*}
Finally we obtain the junction's admittance as
\begin{eqnarray*}
  Y ( \omega ) & = & \frac{\delta  I( \omega  )}{\delta V ( \omega )} =
  \frac{2e}{\hbar} \frac{1}{i \omega} \frac{\delta  I( \omega  )}{\delta
  \varphi ( \omega )}\\
  & = & \frac{\cos   \varphi}{i L_{J }^{\tmop{eff}} \omega}  +2
  \int_{0}^{\infty} \mathd  t S_{I}^{-} ( t ) \frac{e^{i \omega t} -1}{\hbar
  \omega}
\end{eqnarray*}
This expression is a generalized fluctuation-dissipation relation
{\cite{safi_time-dependent_2011}}. Note that the integral contains
contributions from both Cooper pair and quasiparticle tunneling.

\subsection*{Beyond d.c. Josephson effect and Gaussian fluctuations}

Could our mean-field approach be extended to address the full complexity of
the dynamics of Josephson junction? In other words could it handle cases
beyond the restrictions adopted above of (i) static phase difference
({\tmem{i.e.}} strictly dc Josephson effect) and (ii) small
fluctuations/linear response? When lifting restriction (i) the steady-state
analysis conducted above is insufficient, and one needs to replace all
translationally-invariant correlators introduced above by two-time correlators
({\tmem{e.g.}} $S_{H_{T}} ( \tau ) = \langle H_{T} ( \tau ) H_{T} ( 0 )
\rangle \rightarrow S_{H_{T}} ( t,t' ) = \langle H_{T} ( t ) H_{T} ( t' )
\rangle \neq S_{H_{T}} ( t-t' )$ that follow non-Markovian dynamics. In a
situation where the voltage across the JJ is finite and constant on average
(a.c. Josephson effect) these time correlators are cyclostationnary. When the
phase fluctuations become large (ii), because of the non-linear response of
the junction itself the time correlators also become non-Gaussian so that in
Eq. \ref{HTHT} one should distinguish and keep all four correlators of the
charge transfer operator $e^{}$ : $C_{+ +}$, $C_{- -}$, $C_{+ -}$ and $C_{-
+}$. Given the parenthood between the counting fields of Full Counting
Statistics (FCS) and the charge transfer operator $e^{i \hat{\delta}}$
involved here, one could think of adapting/extending FCS results
{\cite{flindt_counting_2010,emary-2011}} to the present problem.


\begin{thebibliography}{100}
  
  
  \bibitem[1]{josephson_possible_1962}\label{bib-josephson_possible_1962}B. D.
  Josephson, \ \href{http://dx.doi.org/10.1016/0031-9163(62)91369-0}{Phys.
  Lett. \tmtextbf{1}, 251 (1962)}.{\newblock}
  
  \bibitem[2]{ambegaokar_tunneling_1963}\label{bib-ambegaokar_tunneling_1963}V.
  Ambegaokar and A. Baratoff,
  \href{http://dx.doi.org/10.1103/PhysRevLett.10.486}{Phys. Rev. Lett.
  {\tmstrong{10}}, 486 (1963)};
  \href{http://dx.doi.org/10.1103/PhysRevLett.11.104}{{\tmstrong{11}}, 104(E)
  (1963)}.{\newblock}
  
  \bibitem[3]{ambegaokar_quantum_1982}\label{bib-ambegaokar_quantum_1982} V.
  Ambegaokar, U. Eckern, and G. Sch{\"o}n,
  \href{http://dx.doi.org/10.1103/PhysRevLett.48.1745}{Phys. Rev. Lett.
  {\tmstrong{48}}, 1745 (1982)}.{\newblock}
  
  \bibitem[4]{schon_quantum_1990}\label{bib-schon_quantum_1990}For a review,
  see Gerd Sch{\"o}n and \ A. D. Zaikin,
  \href{http://dx.doi.org/10.1016/0370-1573(90)90156-V}{Phys. Rep.
  \tmtextbf{198}, 237 (1990)}.{\newblock}
  
  \bibitem[5]{caldeira_quantum_1983}\label{bib-caldeira_quantum_1983}A. O.
  Caldeira and \ A. J. Leggett,
  \href{http://dx.doi.org/10.1016/0003-4916(83)90202-6}{Ann. of Phys. (N.Y.)
  \tmtextbf{149}, 374 (1983)}.
  
  \bibitem[6]{clarke_quantum_1988}\label{bib-clarke_quantum_1988}J. Clarke, A.
  N. Cleland, M. H. Devoret, D. Esteve, and J. M. Martinis,
  \href{http://dx.doi.org/10.1126/science.239.4843.992}{Science
  {\tmstrong{239}}, 992 (1988)}.
  
  \bibitem[7]{wendin_quantum_2007}\label{bib-wendin_quantum_2007}For a review,
  see G. Wendin and V. S. Shumeiko,
  \href{http://dx.doi.org/10.1063/1.2780165}{Low Temp. Phys. {\tmstrong{33}},
  724 (2007)}.
  
  \bibitem[8]{ithier_decoherence_2005}\label{bib-ithier_decoherence_2005}G.
  Ithier, E. Collin, P. Joyez, P. J. Meeson, D. Vion, D. Esteve, F. Chiarello,
  A. Shnirman, Y. Makhlin, J. Schriefl, and \ G. Sch{\"o}n,
  \href{http://dx.doi.org/10.1103/PhysRevB.72.134519}{Phys. Rev. B
  {\tmstrong{72}}, 134519 (2005)}.
  
  \bibitem[9]{martinis_energy_2009}\label{bib-martinis_energy_2009}John M.
  Martinis, M. Ansmann, and \ J. Aumentado,
  \href{http://dx.doi.org/10.1103/PhysRevLett.103.097002}{Phys. Rev. Lett.
  {\tmstrong{103}}, 097002 (2009}).{\newblock}
  
  \bibitem[10]{catelani_quasiparticle_2011}\label{bib-catelani_quasiparticle_2011}G.
  Catelani, J. Koch, L. Frunzio, R. J. Schoelkopf, M. H. Devoret, and \ L. I.
  Glazman, \href{http://dx.doi.org/10.1103/PhysRevLett.106.077002}{Phys. Rev.
  Lett. {\tmstrong{106}}, 077002 (2011)}.
  
  \bibitem[11]{ansari_effect_2012}\label{bib-ansari_effect_2012}M.H. Ansari,
  F.K. Wilhelm, U. Sinha, and A. Sinha,
  \href{http://arXiv.org/abs/1211.4745}{arXiv:1211.4745}.
  
  \bibitem[12]{corlevi_phase-charge_2006}\label{bib-corlevi_phase-charge_2006}S.
  Corlevi, W. Guichard, F. W. J. Hekking, and \ D. B. Haviland,
  \href{http://dx.doi.org/10.1103/PhysRevLett.97.096802}{Phys. Rev. Lett.
  {\tmstrong{97}}, 096802 (2006)}.{\newblock}
  
  \bibitem[13]{pop_measurement_2010}\label{bib-pop_measurement_2010}I. M. Pop,
  I. Protopopov, F. Lecocq, Z. Peng, B. Pannetier, O. Buisson, and \ W.
  Guichard, \href{http://dx.doi.org/10.1038/nphys1697}{Nat. Phys.
  {\tmstrong{6}}, 589 (2010)}.
  
  \bibitem[14]{manucharyan_evidence_2012}\label{bib-manucharyan_evidence_2012}V.E.
  Manucharyan, N.A. Masluk, A. Kamal, J. Koch, L.I. Glazman, and M.H. Devoret,
  \href{http://dx.doi.org/10.1103/PhysRevB.85.024521}{Phys. Rev. {\tmstrong{B
  85}}, 024521 (2012)}.
  
  \bibitem[15]{masluk_microwave_2012}\label{bib-masluk_microwave_2012}N.A.
  Masluk, I.M. Pop, A. Kamal, Z. K. Minev, and M.H. Devoret,
  \href{http://dx.doi.org/10.1103/PhysRevLett.109.137002}{Phys. Rev. Lett.
  {\tmstrong{109}}, 137002 (2012)}.
  
  \bibitem[16]{bell_quantum_2012}\label{bib-bell_quantum_2012}M. T. Bell, I.
  A. Sadovskyy, L. B. Ioffe, A. Yu. Kitaev, and M. E. Gershenson,
  \href{http://dx.doi.org/10.1103/PhysRevLett.109.137003}{Phys. Rev. Lett.
  {\tmstrong{109}}, 137003 (2012)}.
  
  \bibitem[17]{devoret_effect_1990}\label{bib-devoret_effect_1990}M. H.
  Devoret, D. Esteve, H. Grabert, G.-L. Ingold, H. Pothier, and \ C. Urbina,
  \href{http://dx.doi.org/10.1103/PhysRevLett.64.1824}{Phys. Rev. Lett.
  \tmtextbf{64}, 1824 (1990)}.
  
  \bibitem[18]{girvin_quantum_1990}\label{bib-girvin_quantum_1990}S. M.
  Girvin, L. I. Glazman, M. Jonson, D. R. Penn, and \ M. D. Stiles,
  \href{http://dx.doi.org/10.1103/PhysRevLett.64.3183}{Phys. Rev. Lett.
  \tmtextbf{64}, 3183 (1990)}.
  
  \bibitem[19]{ingold_charge_2005}\label{bib-ingold_charge_2005} G.-L. Ingold
  and Y.V. Nazarov, in Single Charge Tunneling, edited by H. Grabert and M.H.
  Devoret, NATO ASI Series B (Plenum, New York, 1992), Vol. 294, pp. 21--107;
  G.-L. Ingold and Y.V. Nazarov,
  \href{http://arXiv.org/abs/cond-mat/0508728}{arXiv:cond-mat/0508728}.
  
  \bibitem[20]{holst_effect_1994}\label{bib-holst_effect_1994}T.~Holst,
  D.~Esteve, C.~Urbina and M. H.~Devoret,
  \href{http://dx.doi.org/10.1103/PhysRevLett.73.3455}{Phys. Rev. Lett.
  {\tmstrong{73}}, 3455 (1994)}.
  
  \bibitem[21]{hofheinz_bright_2011}\label{bib-hofheinz_bright_2011}M.~Hofheinz,
  F.~Portier, Q.~Baudouin, P.~Joyez, D.~Vion, P.~Bertet, P.~Roche and
  D.~Esteve, \href{http://dx.doi.org/10.1103/PhysRevLett.106.217005}{Phys.
  Rev. Lett. {\tmstrong{106}}, 217005 (2011)}.{\newblock}
  
  \bibitem[22]{supplementary_material}\label{bib-supplementary_material}See
  the \href{#Supplementary}{Supplemental Material} for a derivation of Eq.
  (\ref{Y}) and a discussion of possible extensions.
  
  \bibitem[23]{joyez_single-electron_1997}\label{bib-joyez_single-electron_1997}P.
  Joyez and \ D. Esteve,
  \href{http://dx.doi.org/10.1103/PhysRevB.56.1848}{Phys. Rev. {\tmstrong{B
  56}}, 1848 (1997)}.
  
  \bibitem[24]{odintsov_effect_1988}\label{bib-odintsov_effect_1988}A. A.
  Odintsov, Zh. Eksp.Teor. Fiz. 94, 312 (1988)
  [\href{http://www.jetp.ac.ru/cgi-bin/e/index/e/67/6/p1265?a=list}{Sov. Phys.
  JETP 67, 1265 (1988)}.]
  
  \bibitem[25]{note1}\label{bib-note1}The PoET was originally developed for
  tunnel junctions having tunnel resistances much higher than the impedance of
  their environment. In this case the junction behaves nearly as an open
  circuit and phase fluctuations are determined only by the environment.
  
  \bibitem[26]{joyez_how_1998}\label{bib-joyez_how_1998}P. Joyez, D. Esteve
  and M. H. Devoret,
  \href{http://dx.doi.org/10.1103/PhysRevLett.80.1956}{Phys. Rev. Lett.
  \tmtextbf{80}, 1956 (1998}).{\newblock}
  
  \bibitem[27]{goppert_frequency-dependent_1999}\label{bib-goppert_frequency-dependent_1999}G.
  G{\"o}ppert and H. Grabert, C. R. Acad. Sci., Ser. IIb {\tmstrong{327}}, 885
  (1999).
  
  \bibitem[28]{safi_time-dependent_2011}\label{bib-safi_time-dependent_2011}I.~Safi
  and P.~Joyez, \href{http://dx.doi.org/10.1103/PhysRevB.84.205129}{Phys. Rev.
  {\tmstrong{B 84}}, 205129 (2011)}.{\newblock}
  
  \bibitem[29]{grabert_phase_1998}\label{bib-grabert_phase_1998}H. Grabert,
  G.-L. Ingold, and B. Paul,
  \href{http://dx.doi.org/10.1209/epl/i1998-00480-8}{Europhys. Lett.
  {\tmstrong{44}}, 360 (1998)}.
  
  \bibitem[30]{flindt_counting_2010}\label{bib-flindt_counting_2010}C. Flindt,
  T. Novotn{\'y}, A. Braggio, and A.-P. Jauho,
  \href{http://dx.doi.org/10.1103/PhysRevB.82.155407}{Phys. Rev. B 82, 155407
  (2010)}
  
  \bibitem[31]{emary-2011}\label{bib-emary-2011}C. Emary,
  \href{http://dx.doi.org/10.1088/0953-8984/23/2/025304}{J. Phys.: Condens.
  Matter {\tmstrong{23}} 025304 (2011)}.
  
  
\end{thebibliography}
\end{document}